\documentclass[osajnl,onecolumn,showpacs,superscriptaddress,10pt]{revtex4-1}
\usepackage{amsmath,amssymb,graphicx,placeins}
\begin{document}

\title{
						Compression of Ultrashort Laser Pulses
				via Gated Multiphoton Intrapulse Interference Phase Scans
			}

\author{Alberto Comin}
	\email{Corresponding author: alberto.comin@cup.lmu-muenchen.de}
\author{Richard Ciesielski}
\author{Giovanni Piredda}
\author{Kevin Donkers}
\author{Achim Hartschuh}
\affiliation{Department Chemie \& CeNS, LMU M\"unchen, 81377 M\"unchen, Germany}

\begin{abstract}
Delivering femtosecond laser light in the focal plane of a high numerical aperture microscope objective is still a challenge, despite significant developments in the generation of ultrashort pulses. One of the most popular techniques, used to correct phase distortions resulting from propagation through transparent media, is the multiphoton intrapulse interference phase scan (MIIPS). The accuracy of MIIPS however is limited when higher order phase distortions are present. Here we introduce an improvement, called Gated-MIIPS, which avoids shortcomings of MIIPS, reduces the influence of higher order phase terms, and can be used to more efficiently compress broad band laser pulses even with a simple 4f pulse shaper setup. In this work we present analytical formulas for MIIPS and Gated-MIIPS valid for chirped Gaussian pulses; we show an approximated analytic expression for Gated-MIIPS valid for arbitrary pulse shapes; finally we demonstrate the increased accuracy of Gated-MIIPS via experiment and numerical simulation.
\\ \\
This paper was published in Journal of the Optical Society of America B and is made available as an electronic reprint with the permission of OSA. The paper can be found at the following URL on the OSA website:  \href{url}{http://www.opticsinfobase.org}. Systematic or multiple reproduction or distribution to multiple locations via electronic or other means is prohibited and is subject to penalties under law.

\end{abstract}

\ocis{(190.7110) Ultrafast nonlinear optics; (300.6530) Spectroscopy, ultrafast; 
(320.5540) Pulse shaping.}

\maketitle 

\section{Introduction}

Short laser pulses are widely used in microscopy, spectroscopy and micro-machining.\cite{Diels2006} They allow the study of elementary processes in real time and, thanks to their high peak power, they are ideal for investigating nonlinear phenomena.
However they are difficult to handle, since as they propagate through any medium, including air, they collect phase distortions. This is particularly critical in the case of focusing through high numerical aperture objectives, since the amount of glass contained in such devices is enough to broaden the pulse duration of a femtosecond pulse by several orders of magnitude.\cite{Muller1998,Accanto2014}
There are several techniques which can in principle be used to compensate phase distortions and deliver transform limited pulses to the sample plane  of a microscope. The most straightforward are based on prism and grating compressors, which can be used to minimize the quadratic and cubic terms in the phase expansion.\cite{Diels2006}

Pulse shaping has, in principle, the ability to compensate arbitrary phase distortions,\cite{Weiner2011} but how to determine which phase to program in the pulse shaper is  less straightforward. Genetic algorithms, for instance, require substantial measurement time and computational effort and do not always yield the desired pulse phase profile in a reproducible way. Frequency-resolved optical gating (FROG) is perhaps one of the most popular tools to retrieve to phase of femotosecond pulses,\cite{Fittinghoff1998} and there are several others ingenious techniques.\cite{Iaconis1998,Loriot2013} A significant step towards reliable and reproducible compression of ultra-short laser pulses was achieved  by the group of Dantus, with the development of MIIPS.\cite{Xu2006a} The authors demonstrated the possibility to compensate the phase distortion introduced by a  $0.60 {\rm NA}$ objective by only using MIIPS.\cite{Xu2006} However the compensation of higher NA objectives benefits from external  compression (for instance by a prism pair) to retain the accessible phase range of the shaper.\cite{Xu2006}

\begin{figure}[htb]
\centering\includegraphics[width=8.4cm]{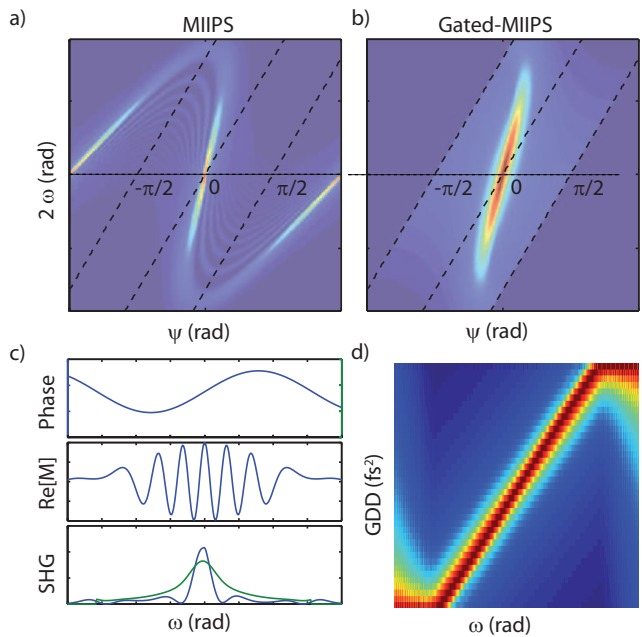}
\caption{Multiphoton intrapulse interference phase scans: (a) simulated Miips trace of a femtosecond laser pulse with large cubic phase distortion. (b) equivalent Gated-MIIPS scan (c) from the top: phase modulation used in the MIIPS scan; real part of the phase-amplitude modulation used in Gated-MIIPS; SHG spectrum of the modulated femtosecond pulse for MIIPS (blue curve) and Gated-MIIPS (green curve, multiplied by 4). (d) group delay dispersion obtained geometrically by skewing and rescaling the Gated-MIIPS trace. }
\label{fig:GatedMiips}
\end{figure}

In this paper we show that MIIPS can be improved by complementing it with an amplitude gate, that is scanned across the spectrum. We demonstrate that this gated version of MIIPS (Fig.\ref{fig:GatedMiips}) estimates more accurately the spectral phase, without the need to increase the measurement time. It is also less affected by systematic errors originating from higher order phase distorsions and shaper artifacts.

The work is organized as follows: in section \ref{sec:Miipsintro} we give a brief introduction to the MIIPS technique; in section \ref{sec:Miipschallenges} we discuss the limitations of the standard implementation of Miips; in section \ref{sec:GatedMiips} we introduce Gated-Miips, in section \ref{sec:analytical} we derive analytical models for MIIPS and Gated-MIIPS; in section \ref{sec:NumSimulation} we illustrate, by numerical simulation, the advantages of Gated-MIIPS and, in section \ref{sec:ExpData}, we show experimental data indicating that Gated-MIIPS is better suitable than standard MIIPS in estimating the phase of a femtosecond laser pulse.
\begin{table}[h!]
  \caption{List of symbols used in this article.}
	\label{tab:ListOfAbbreviations}
  \begin{center}
    \begin{tabular}{c | c | c}
    \hline
    Symbol & Unit &Meaning \\
    \hline
		$\omega$ & rad/fs & angular frequency\\
		$\omega_0$ & rad/fs & pulse central angular frequency\\
		$\Delta \omega$ & rad/fs & pulse width (at 1/e)\\
    $E(\omega)$ & V/m & complex electric field\\
		$\phi(\omega)$ & rad & electric field phase\\
		$\Phi_0$ & rad & phase modulation amplitude\\
		$\tau$ & fs & modulation frequency\\
		$\psi$ & rad & modulation phase\\
		$\psi_m(\omega)$ & rad & $\psi$ value that maximize ${\rm SHG}(2 \omega)$\\
		$\Delta\psi$ & rad & step variation of $\psi$\\
		$\epsilon$ & dimensionless & relative error of GDD\\
		$\Delta t$ & fs & temporal FWHM\\
		$\sigma$ & rad &amplitude gate width (at 1/e)\\
    \hline
    \end{tabular}
  \end{center}
\end{table}

\section{Multiphoton Intrapulse Interference Phase}
\label{sec:Miipsintro}

MIIPS was designed as a tool to control multi-photon processes and to characterize and compress ultrashort laser pulses.\cite{Dantus2004} It employs a pulse-shaper setup, in order  to estimate the group delay dispersion (GDD) and to compensate arbitrary complex spectral phase distortions. The main concept is that the GDD is correlated with the intensity of Second Harmonic Generation (SHG): at any given frequency the SHG is maximized if the GDD is null.\cite{Walowicz2002} It follows that a transform limited (flat phase) pulse can be obtained by applying suitable phase masks and maximizing the SHG for the whole spectrum.\cite{Walowicz2002} This can be seen from the equation for the  SHG at the frequency $\omega$ in the ideal case (assuming $\chi^2=1$) as: 
\begin{equation} 
\label{eq:SHG}
	{\rm SHG}(2 \omega) = \left| \int_{-\infty}^{+\infty}
	\left| E(\omega-\Omega) \right| \left| E(\omega+\Omega) \right| \right. 
	 \exp\left[ i (\phi(\omega-\Omega)+\phi(\omega+\Omega) \right]
	{\rm d}\Omega \left.\vphantom{\int} \right|^2
\end{equation}
Where $E(\omega)\exp(i \phi)$ is the amplitude of the electric field at angular frequency $\omega$. A list of the symbols used in this articles can be found in Table \ref{tab:ListOfAbbreviations}. Since the first two factors of the integrand  in Eq.\ref{eq:SHG} are positive, the SHG is maximized when the argument of the exponential is zero. Expanding the phase in series and retaining the second order, the approximated  equation for the SHG becomes:
\begin{equation}
\label{eq:SHGapprox}
	{\rm SHG}(2 \omega) = 
	\left| \int_{-\infty}^{+\infty} 
	\left| E(\omega-\Omega) \right| \left| E(\omega+\Omega) \right|
	\exp\left( i \frac{\partial^2\phi(\omega) }{ \partial\omega^2} \Omega^2 \right)
	{\rm d}\Omega  \right|^2
\end{equation}

It follows that the SHG is maximized when the group delay dispersion
${\rm GDD } = \ddot{\phi}(\omega) $ is equal to zero. This condition is at the basis of MIIPS, and it is valid as long as the phase can  be locally approximated by a second order polynomial. The standard implementation of MIIPS involves adding a sinusoidal modulation to the phase of the laser pulse, typically using a pulse-shaper:
\begin{equation} 
f(\omega) = \exp \left\{ i \Phi_0 \,
 \sin\left[ \tau \, (\omega-\omega_0)-\psi \right] \right\}
\label{eq:MiipsTestPhase}
\end{equation}
We note that such a sinusoidal phase modulation can also be seen, in the time domain, as the generation of a pulse train.\cite{Hacker2001,Lozovoy2005} The modulation frequency $\tau$, which has the units of time, corresponds to the temporal separation between subsequent pulses.\cite{Hacker2001}
The GDD is then estimated by varying the modulation phase $\psi$ (Fig.\ref{fig:GatedMiips}) and determining for which values the SHG intensity is locally maximized. 
 Combining Eq.\ref{eq:SHG} and Eq.\ref{eq:MiipsTestPhase}  and using the prosthaphaeresis formulae, it is possible to show that the SHG of a laser pulse subject to sinusoidal phase modulation is maximized when:
\begin{equation}
\phi(\omega-\Omega)+\phi(\omega+\Omega)+2 \Phi_0
\cos (\tau \Omega) \sin \left[ (\omega-\omega_0) \tau - \psi \right] = 0
\label{eq:MiipsCondition}
\end{equation}
which means that the GDD can be estimated, to the second order of accuracy, as:
\begin{equation} 
	\ddot{\phi}(\omega) = \Phi_0\,\tau^2 \, \sin\left[ \tau  \, (\omega-\omega_0)-\psi_m(\omega) \right] 
\end{equation}
Where $\psi_m(\omega)$ is the value of the scanning parameter $\psi$ which maximizes the SHG for the frequency $\omega$. The phase is then recovered numerically by double integration.
For a more extensive description of the implementation of MIIPS and its applications we refer to the works by Dantus, quoted in the references. We note that the zero and first order terms in the Taylor  expansion of the phase do not contribute to second harmonic generation (Eq.\ref{eq:SHG}) and therefore are not measured by MIIPS.\cite{Xu2006} The zero order is responsible for  the  carrier-envelope phase and the first order leads to a translation of the pulse along the time axes.

\section{Challenges of MIIPS}
\label{sec:Miipschallenges}

A known limitation of MIIPS is that it only considers the $2^{\rm nd}$ order correction to the phase at a given frequency, neglecting higher orders in the Taylor expansion.\cite{Xu2006,Lozovoy2004} Although this can lead to tolerable errors for not overly distorted pulses, in other cases it can significantly reduce the accuracy of MIIPS as discussed below. Xu et al. showed that the systematic error is of the order of $(\tau/\Delta t)^2/12$, where $\tau$ is the MIIPS modulation frequency and $\Delta t$ is the transform limited pulse duration.\cite{Xu2006} From the above estimate, it might seem that accurate results could always be obtained for very short pulse duration, just by choosing a very small modulation frequency, and indeed an ideal MIIPS simulation, which does not take into account any of the shaper artifacts, would confirm this conclusion.

\begin{figure}[htb] 

\centering\includegraphics[width=8.4cm]{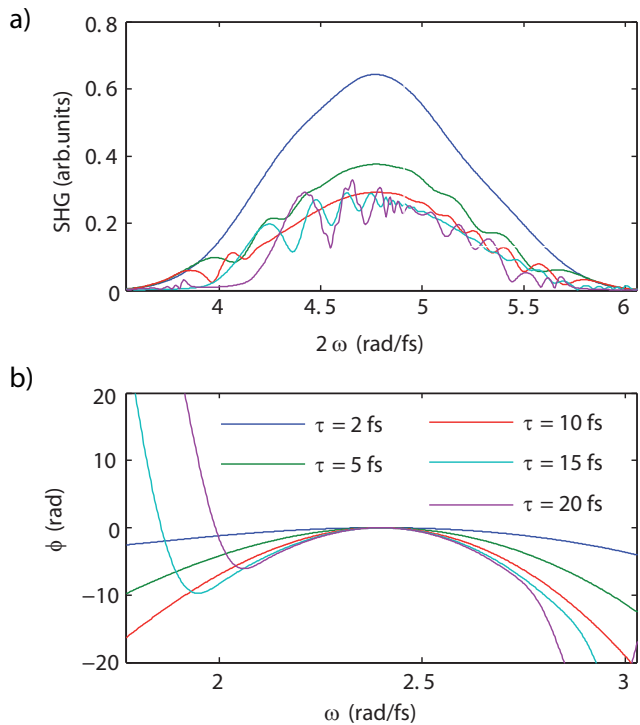}
\caption{Performance of MIIPS for scans obtained for a $\Delta t = 5.3 {\rm fs}$ laser pulse after propagation through $10{\rm cm}$ glass, using different phase modulation frequencies and keeping the maximum correction constant  $\Phi_0 \tau^2 = 2500{\rm fs}^2$.  (a) SHG uncorrected and MIIPS corrected spectra for different phase modulation frequencies. (b) Residual phase for the uncorrected laser pulse (demagnified $10^4$ times) and after  MIIPS correction using different phase modulation frequencies.}
\label{fig:MiipsPerformance}
\end{figure}
In Fig.\ref{fig:MiipsPerformance} we report several simulated MIIPS scans performed on a $5.3{\rm fs}$ laser pulse centered at $\omega_0=2.4{\rm rad}/{\rm fs}$ ($785 {\rm nm}$), after propagation through $10{\rm cm}$ of BK7 glass, which mimic the optical path length of a microscope objective. For each scan we report the residual spectral phase and the SHG spectrum. In order to compare the different modulation parameters we kept the maximum GDD correction constant (see Eq.\ref{eq:maxGDD}). As expected, the best performance in terms of maximum SHG and residual phase are  obtained for the smaller values of the modulation frequency $\tau$.

In practice the adjustment of the modulation parameters is subtler because, for highly distorted ultrashort pulses, it  requires setting the modulation amplitude to very high values, which in turn deteriorates the spectrum by introducing shaper artifacts, like for instance diffraction caused by phase grating and ripples in transmitted spectra. The former can be minimized by employing a double pass 4f pulse shaper setup.\cite{Brinks2011} The latter can be explained as follows.

By definition the spectral phase is given by multiples of $2\pi$ radians. However due to cross-talk between the SLM pixels, phase and amplitude distortions can be observed in the spectral regions where the phase crosses $2\pi$, known as wrapping.\cite{Vaughan2006} Therefore, if the amount of wrapping is very high, the spectrum transmitted through the SLM can be too distorted to be accurately characterized. This poses a limitation to the total phase which can be compensated and it is particularly severe for very short pulses ($<10 {\rm fs}$) in combination with high-numerical aperture objectives.
\begin{figure}[htb] 
\label{fig:MiipsTrajectory}
\centering\includegraphics[width=8.4cm]{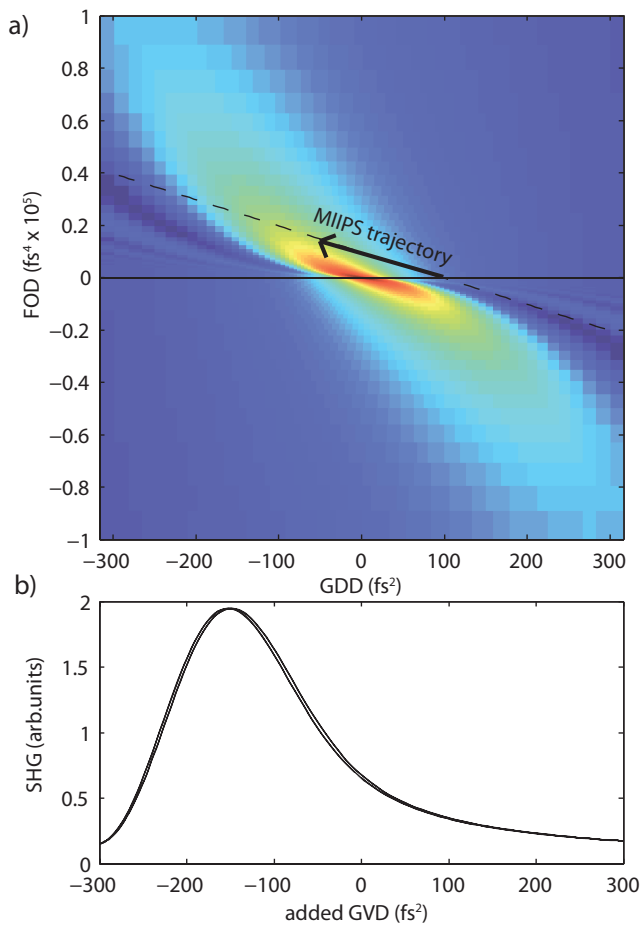}
\caption{Phase dependence of the SHG intensity at the central frequency of a 100nm broad laser pulse centered at 800nm (a) Map of the SHG in function of the second and fourth order phase terms. The dashed line represents the points explored by a typical MIIPS trace. (b) The maximum SHG intensity along the MIIPS trajectory does not correspond to zero GDD.}
\end{figure}

For the above reasons, a common choice is to set the phase modulation frequency $\tau$ equal to the transform limited pulse duration $\Delta t$.\cite{Xu2006} This is a good compromise because lower modulation frequency would result in having a too high modulation amplitude, which would cause phase-wrapping artifacts, as we previously discussed. Higher modulation frequencies, on the other hand, are not optimal because the second order polynomial expansion in Eq.\ref{eq:SHGapprox} loses validity.
Indeed, since MIIPS uses a sinusoidal phase mask, it introduces both second (SOD) and fourth order (FOD) corrections, linked by the relation: ${\rm FOD}= - {\rm SOD} \cdot \tau^2$. In a 2D map which represents the SHG intensity as a function of GDD and FOD values (Fig.\ref{fig:MiipsTrajectory}), MIIPS maximizes the signal along a linear trajectory and from Fig.\ref{fig:MiipsTrajectory} it can be seen that this maximum does not necessarily correspond to a zero GDD. 

\subsection*{MIIPS Resolution} 
For a defined set of modulation parameters, the maximum GDD which can be compensated by a single MIIPS iteration is:
\begin{equation} 
\label{eq:maxGDD}
{\rm GDD_{max}} = \Phi_0 \, \tau^2
\end{equation}
On the other hand, the minimum GDD which can be compensated depends on the size of the phase steps and on the maximum relative error which can be accepted for the GDD. If $\Delta \psi$ is the increment of the argument of the sinusoidal modulation, and $\epsilon = \Delta {\rm GDD}/{\rm GDD}$ is the maximum relative error on the GDD, then it follows that:
\begin{equation} 
\label{eq:GDDrelerr}
\left|{\rm GDD_{min}}\right| = \Phi_0 \, \tau^2 \,
\left(
	1 + \left( \frac {\epsilon}{\Delta \psi} \right)^2
\right)^{-1/2}
\end{equation}

The dynamical range, defined as the ratio of the maximum and the minimum GDD is therefore:
\begin{equation} 
	D = \left(
	1 + \left( \frac{\epsilon}{\Delta \psi} \right)^2
	\right)^{1/2} \approx \frac{\epsilon}{\Delta \psi}
\end{equation}
Which leads, for a scan of 1000 points and an accuracy of $10\%$, to a dynamical range of only 16. This explains why MIIPS needs to be iterated multiple times to compensate the phase of laser pulses.
This limitation can be mitigated by limiting the spectral bandwidth (and the range of GDD values) around the frequency at which the GDD is being compensated, as done in Gated-MIIPS discussed below.


\section{Gated-MIIPS}
\label{sec:GatedMiips}

We developed Gated-MIIPS to avoid the limitations of MIIPS when compensating large phase distortions caused by broad spectra and high numerical aperture objective. The idea, illustrated in Fig.\ref{fig:GatedMiips}, is to enable 
using higher modulation frequencies for the sinusoidal phase, while avoiding systematic errors due to higher order phase terms. In the next sections, however, we will show that Gated-MIIPS is more accurate that MIIPS also with the same choice of modulation parameters. 

Gated-MIIPS can be readily implemented using a pulse shaper which provides both phase and amplitude modulation. The amplitude modulation is exploited to gate the spectrum around a specific frequency (Fig.\ref{fig:GatedMiips}c), therefore improving the validity of the second order polynomial expansion of the phase, even in the case of significant higher order contributions, as will be shown below. It then becomes possible to use higher values of the modulation frequency $\tau$ and lower values of the modulation amplitude $\Phi_0$ minimizing the phase wrapping. In the case of a Gaussian gate the modulation that is applied by the pulse shaper can be written as a function of the phase terms:
\begin{equation}
\begin{split}
M(\omega) = 
\exp \left\{ - \left[
\frac{\tau \, (\omega-\omega_0)-\psi} {\sigma}
\right]^2  + i \Phi_0 \sin \left[\tau \, (\omega-\omega_0)-\psi \right]  \right\}
\end{split}
\label{eq:gate}
\end{equation}

 As it can be seen from Eq.\ref{eq:gate}, in Gated-MIIPS a Gaussian amplitude mask of width $\sigma$ (here called gate) is translated alongside the phase modulation, using the scanning parameter $\psi$.
To improve the accuracy for the GDD values closer to the gate boundaries ($\pm \Phi_0 \tau^2$) the Gated-MIIPS signal $\mathcal{G}(2 \omega)$ is then obtained by dividing the SHG spectrum by the square of the Gaussian amplitude mask.
\begin{equation}
\mathcal{G}(2 \omega) =
{\rm SHG}(2 \omega) \cdot
 \exp \left\{
	4\left[ \frac{\tau \, (\omega-\omega_0)-\psi}{\sigma}\right]^2
\right\}
	\label{eq:GatedMiipssignal}
\end{equation}

By combining Eq.\ref{eq:SHG} with Eq.\ref{eq:gate}, in analogy with what previously discussed for the standard MIIPS case, it can be shown that:

\begin{equation} 
\begin{split}
\mathcal{G}(2 \omega) &= 
\left|\int_{-\infty}^{+\infty}
\left| E(\omega-\Omega) \right|
\left| E(\omega+\Omega) \right|\cdot
\exp \left(
	-\frac{2 \tau^2 \Omega^2}{\sigma^2}
\right) \cdot  \right. \\
&\exp \left\{\right. i \left[\right.
\phi(\omega-\Omega)+\phi(\omega+\Omega)+
\left. 2 \Phi_0 \cos (\tau \Omega) \sin \left[\tau (\omega-\omega_0) - \psi\right]
\left.\right]
\left.\right\}
{\rm d}\Omega \vphantom{\int}\right|^2
\end{split}
\label{eq:GatedMiips}
\end{equation}

By direct inspection, it can be seen from Eq.\ref{eq:GatedMiips} that the Gated-MIIPS signal $\mathcal{G}(2 \omega)$ is maximized by the same condition expressed by the Eq.\ref{eq:MiipsCondition}.
Therefore, in the ideal case Gated-MIIPS should provide the same results as the standard MIIPS. The advantage of Gated-MIIPS stems from the fact that the presence of the Gaussian term in Eq.\ref{eq:GatedMiips} improves the validity of the second order expansion of the spectral phase (Eq.\ref{eq:SHGapprox}).

In analogy with standard MIIPS,\cite{Lozovoy2005} the Gated-MIIPS signal can also be expressed analytically for a few simple pulse shapes. For instance, analytical expressions for the SHG signal of a transform limited Gaussian pulse, subject to sinusoidal phase modulation, have been given by Hacker et al.\cite{Hacker2001} and reproposed by Lozovoy et al.\cite{Lozovoy2005}.  
However, to the best of our knowledge, an analytical expression of the MIIPS signal in the case of a chirped Gaussian pulse has not been shown yet. In the next section we will derive it for the cases of both MIIPS and Gated-MIIPS. Such an expression can be useful for fitting a MIIPS trace without relying on a peak finding algorithm, as it is usually done.\cite{Xu2006}  Additionally, we will show that an approximated analytic expression for Gated-MIIPS signal can be also be written for a generic pulse shape, provided that the field amplitude varies slowly with respect to the Gaussian gate (Eq.\ref{eq:gate}).

\section{Analytical Models for MIIPS and Gated-MIIPS}
\label{sec:analytical}

\subsection{MIIPS on a chirped Gaussian Pulse}

Let us assume a Gaussian pulse given by:
\begin{equation}
\label{eq:ChirpedPulse}
	E(\omega) = \exp \left(
	- \frac{(\omega-\omega_0)^2} {{\Delta\omega}^2}
	+ i \, \phi(\omega)
	\right)
\end{equation}
By substituting Eq.\ref{eq:ChirpedPulse} into Eq.\ref{eq:SHG} we obtain:

\begin{equation}
\label{eq:MiipsGaussianPulse}
\begin{split}
{\rm SHG}(2 \omega) &= 
\exp \left\{ -4 \, \frac { \left( \omega-\omega0 \right) ^2}{{
\Delta \omega }^2}\right\} \cdot \\
 &\left|  
 \int _{-\infty }^{\infty }\!
\exp \left\{-2\,\frac {{\Omega}^2} {{\Delta \omega }^2} 
+i\, \left[ \phi(\omega-\Omega)+\phi(\omega+\Omega)+  2\Phi_0\,\cos \left( \tau\,\Omega \right)
 \sin \left(\tau \, (\omega-\omega_0)  -
\psi \right)\right] \vphantom{\int} \right\}
{{\rm d}\Omega} \right|^{2}
\end{split}
\end{equation}
Now if, as in standard MIIPS, we approximate the phase $\phi(\omega)$ by a second order polynomial, it becomes possible to calculate the integral by using the Jacobi-Anger expansion, given by the formula:
\begin{equation}
\label{eq:JacobiAnger}
	\exp \left( i \, z  \cos (\theta) \,  \right) = 
	\sum_{n=-\infty}^{+\infty} \left(
	i^n J_n(z) \exp(i \, n \, \theta)
	\right)
\end{equation}
Where $J_n(z)$ is the Bessel function of the first kind. The resulting SHG is given by:
\begin{equation}
\label{eq:MiipsFullAnalytical}
\begin{split}
{\rm SHG}(2 \omega) = 
\frac
	{\pi {\Delta\omega}^2}
	{\sqrt{ {\Delta\omega}^4 {\ddot{\phi}(\omega)}^2 + 4 }}
	\exp \left(-
	\frac
		{4 (\omega-\omega_0)^2}
		{{\Delta\omega}^2}
	\right) \cdot 
\sum_{n=-\infty}^{+\infty}
{
	\exp \left(-
	\frac{ {\Delta\omega}^2 \tau^2\, n^2 }
	{{\Delta\omega}^4 {\ddot{\phi}(\omega)}^2+4}
	\right)
	{J_n\left(
	2 \Phi_0 \sin \left(
		\tau (\omega-\omega_0)-\psi
	\right)
	\right)}^2
}
\end{split}
\end{equation}
 Eq.\ref{eq:MiipsFullAnalytical} can be accurately computed using a limited number of terms, because as $\left| n \right|$ increases the contribution of each added term decreases exponentially. A simpler formula, accurate to second order in the phase, can be obtained by doing the Taylor expansion of the phase term in Eq.\ref{eq:MiipsGaussianPulse}.

After some algebraic manipulation one obtains:
\begin{equation}
\begin{split}
{\rm SHG}(2 \omega,\phi) =  \pi {\Delta\omega}^2
\exp \left(
 -\frac{4(\omega-\omega_0)^2}{{\Delta\omega}^2}
\right)  
 \left\{
4+{\Delta\omega}^4
\left[
\,\ddot{\phi}(\omega)- \tau^2 \Phi_0  \sin (\tau (\omega-\omega_0) -\psi)
\right]^2
\right\}^{-1/2}
\end{split}
\label{eq:MiipsApproxAnalytical}
\end{equation}
By differentiating the above formula with respect to $\psi$ and letting ${\partial\, {\rm SHG}(2 \omega)}/{\partial\psi}=0$, we find the explicit expression of the GDD in function of the scanning parameter $\psi$, which was already found in Eq.\ref{eq:MiipsCondition}.

\subsection{Analytical expression for Gated-MIIPS}

For the specific case of a Gaussian pulse, the analytical expression for the Gated-MIIPS signal is very similar to the one we just derived for MIIPS. By substituting Eq.\ref{eq:ChirpedPulse} into Eq.\ref{eq:GatedMiips}, after some algebraic manipulation, one arrives at a formula similar to Eq.\ref{eq:MiipsGaussianPulse}, except for the substitution ${\Delta\omega}^2 \rightarrow L^2=({\Delta\omega}^2 \sigma^2)/(\sigma^2+{\Delta\omega}^2 \tau^2)$ inside the integration sign.
\begin{equation}
\label{eq:GatedMiipsGaussianPulse}
\begin{split}
\mathcal{G}(2 \omega) &= 
\exp \left\{ -4 \, \frac { \left( \omega-\omega0 \right) ^2}{{
\Delta \omega }^2}\right\} \cdot  \\
& \left|  
 \int _{-\infty }^{\infty }\!
\exp \left\{-2\,\frac {{\Omega}^2} {L^2} 
+i\, \left[ \phi(\omega-\Omega)+\phi(\omega+\Omega)+   2\Phi_0\,\cos \left( \tau\,\Omega \right) \sin \left[ \tau\,(\omega-\omega_0) -
\psi \right]\right] \vphantom{\int} \right\}
{{\rm d}\Omega} \right|^{2}
\end{split}
\end{equation}
We note that, since it is always $L<\Delta\omega$, Gated-MIIPS traces are normally ticker that MIIPS traces recorded with the same modulation parameters (see Fig).

Using the Jacobi-Anger expansion is then straightforward to derive the analog of Eq.\ref{eq:MiipsApproxAnalytical}.
In the same way, the approximated Gated-MIIPS signal of a chirped Gaussian pulse is given by expressions equivalent to Eq.\ref{eq:MiipsApproxAnalytical}:
\begin{equation}
\begin{split}
\mathcal{G}(2 \omega,\phi) =  \pi {L}^2
\exp \left(
 -\frac{4(\omega-\omega_0)^2}{{\Delta\omega}^2}
\right)  
 \left\{
4+{L}^4
\left[
\,\ddot{\phi}(\omega)- \tau^2 \Phi_0  \sin (\tau (\omega-\omega_0) -\psi)
\right]^2
\right\}^{-1/2}
\end{split}
\label{eq:GatedMiipsApproxAnalytical}
\end{equation}
This confirms that, under this approximation, Gated-MIIPS provides the same information as MIIPS. In the next section, however, we will show by numerical simulation that, in realistic conditions where higher order phase terms cannot be neglected, Gated-MIIPS provides a significant advantage over the standard implementation of MIIPS.

A more general formula can be derived for an arbitrary shaped pulse, provided that its amplitude vary slowly with respect to the width  of the gate $\sigma$. By expanding in series both the phase term the field amplitude in Eq.\ref{eq:GatedMiips} one can obtain:

\begin{equation}
\label{eq:GatedMiipsGeneral}
\begin{split}
\mathcal{G}(2 \omega) &= 
\frac{\pi \sigma^2}{4} \left\{
\left[
		\ddot{\phi}(\omega) - \tau^2 \Phi_0 \sin (\tau (\omega-\omega_0)  - \psi)
	\right]^2 
\sigma^4 + 4 \tau^4 \right\}^{-3/2} \cdot \\
 &\left\{
4 \sigma^4 \left|E(\omega)\right|^4
\left[
		\ddot{\phi}(\omega) - \tau^2 \Phi_0 \sin (\tau (\omega-\omega_0)  - \psi)
	\right]^2 
  +  \left[
\sigma^2 \left(
\left|E(\omega)\right| \, 
\frac{\partial^2\left|E(\omega)\right|}{\partial\omega^2}
 - \left( 
\frac{\partial \left|E(\omega)\right|}{\partial\omega}
 \right)^2
\right)+
4 \tau^2 \left|E(\omega)\right|^2 
\right]^2
\right\}
\end{split}
\end{equation}

\section{Numerical Comparison}
\label{sec:NumSimulation}

\begin{figure}[htb] 
\centering\includegraphics[width=8.4cm]{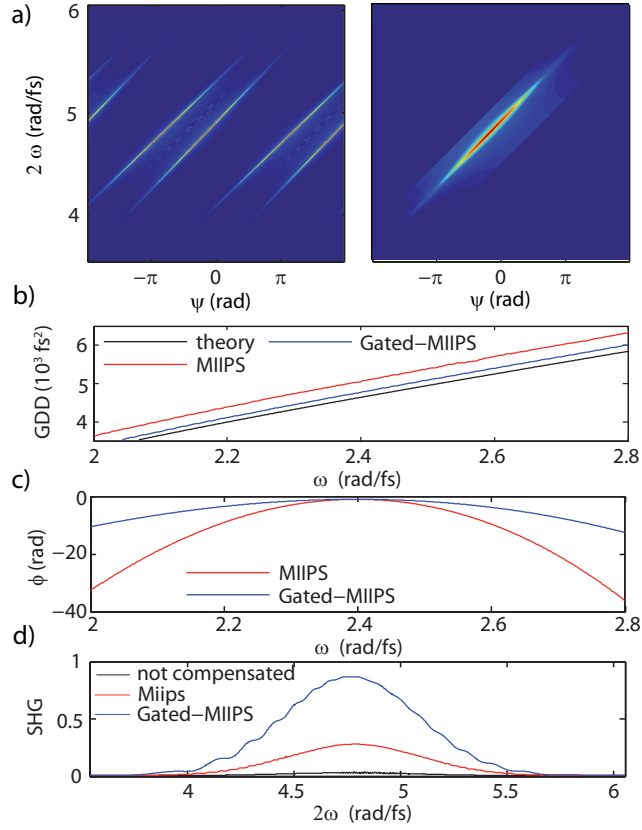}
\caption{Simulation of the compensation of a $5.3\,{\rm fs}$
laser pulse centered at $2.4\,\textrm{rad/fs}$ after propagation through $10\, \textrm{cm}$ of BK7 glass.  (a) standard MIIPS (left) and Gated-MIIPS (right) maps, obtained with $\Phi_0=100\, {\rm rad}$, $\tau=10\,\textrm{fs}$ and $\sigma=0.5\,\textrm{rad}$. (b) GDD measured by a single  iteration of MIIPS (red line) and Gated-MIIPS (blue line), together with actual GDD value (black line). (c)  residual phase after a single  iteration of MIIPS (red line) and Gated-MIIPS (blue line). (d) SHG signal of uncompensated SHG (black line) and after a single  iteration of MIIPS (red line) and Gated-MIIPS. (blue line).}
\label{fig:glasscompression10cm}
\end{figure}

In order to test the efficacy of the Gated-MIIPS we performed a series of simulated measurements (see Fig.\ref{fig:glasscompression10cm}a-b), in which we compared the ability of MIIPS and Gated-MIIPS in correctly estimating the phase introduced by $10\,\textrm{cm}$ of glass on a $5.3\,\textrm{fs}$ laser pulse centered at $\omega_0 = 2.4 \, \textrm{rad/fs}$. The modulation parameters were $\Phi_0=100\, {\rm rad}$ and $\tau=10\,\textrm{fs}$, and the gate width was $\sigma=0.5\,\textrm{rad}$.


In Fig.\ref{fig:glasscompression10cm}c-d we report, for the two cases, the actual and estimated values of GDD, the SHG spectrum before and after phase correction obtained by a single iteration, the estimated phase and the residual phase.  From the traces reported in Fig.\ref{fig:glasscompression10cm} it can be seen that the Gated version of MIIPS is superior in both estimating the group velocity dispersion and yielding greater SHG after one iteration. The difference could be even greater if we consider that, to compensate such a considerable GDD, we had to employ a modulation amplitude $\Phi_0$ of $100\, \textrm{rad}$, which could cause significant phase-wrapping distortions. In order to reduce phase wrapping, we would have to decrease $\Phi_0$ and correspondingly increase $\tau$ (see Eq.\ref{eq:maxGDD}). This would reduce the performance of MIIPS, because of the increased contribution from higher order phase terms, as discussed above. Gated-MIIPS, however, would be relatively unaffected.

\begin{figure}
\centering\includegraphics[width=8.4cm]{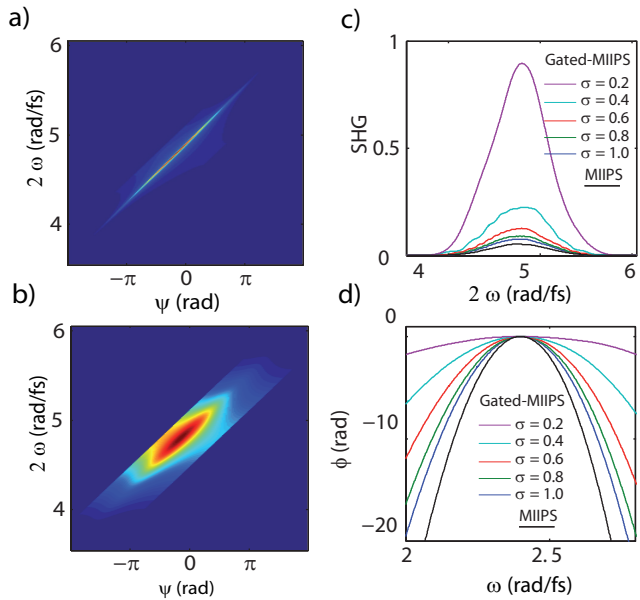}
\caption{Gated-MIIPS for different settings of the gate width, simulated for a $5.3\,{\rm fs}$ laser pulse centered at $2.4\,\textrm{rad/fs}$ after propagation through $10\, \textrm{cm}$ of BK7 glass. For all cases the modulation parameters were $\Phi_0=100\ {\rm rad}$ and $\tau=10\,\textrm{fs}$. Standard MIIPS is also shown for comparison.(a) Gated-MIIPS map corresponding to $\sigma=1\,\textrm{rad}$ (b) Gated-MIIPS map corresponding to $\sigma=0.2\,\textrm{rad}$ (c) SHG after a single iteration of Gated-MIIPS for $\sigma$ varying between $1$ and $0.2$ (d) Residual phase after a single iteration of Gated-MIIPS for $\sigma$ varying between $1$ and $0.2$}
\label{fig:GateDependence}
\end{figure}

One important parameter that needs to be considered when comparing MIIPS with Gated-MIIPS is the choice ofthe gate width $\sigma$ with respect to the pulse bandwidth $\Delta\Omega$ modulation frequency $\tau$. To some extent narrower gate values yield better results, as shown in Fig. \ref{fig:GateDependence}). However, reducing the gate also means decreasing the intensity of the SHG spectra recorded during the Gated-MIIPS scan. Therefore the integration time needs to be adjusted accordingly. The ideal choice of the gate width is then a compromise between phase accuracy and measurement time.

\begin{figure}[htb] 
\centering\includegraphics[width=8.4cm]{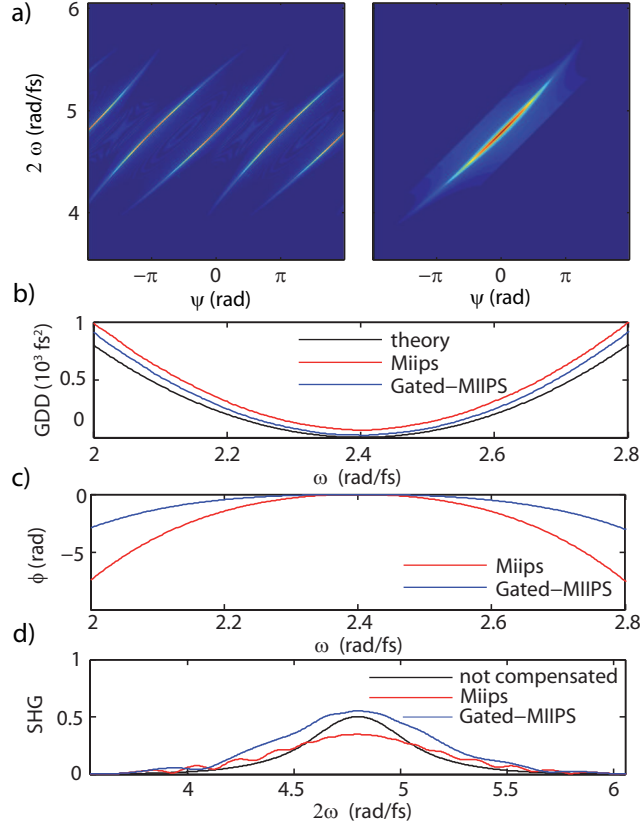}
\caption{Simulation of the compensation of a $5.3\,{\rm fs}$
laser pulse centered at $2.4\,\textrm{rad/fs}$ with significant ($10^4\,{\rm fs}^4$) fourth order phase distortion.  (a) standard MIIPS (left) and Gated-MIIPS (right) maps, obtained with $\Phi_0=20\, {\rm rad}$ and $\tau=10\,\textrm{fs}$. (b) GDD measured by a single  iteration of MIIPS (red line) and Gated-MIIPS (blue line), together with actual GDD value (black line). (c)  residual phase after a single  iteration of MIIPS (red line) and Gated-MIIPS (blue line). (d) SHG signal of uncompensated SHG (black line) and after a single  iteration of MIIPS (red line) and Gated-MIIPS (blue line). Standard MIIPS in this case reduces SHG.} 
\label{fig:FourthOrder}
\end{figure}


As previously discussed, the reason of the success of Gated-MIIPS in compensating the phase introduced by propagation through transparent media lies in the ability of dealing with higher order phase contributions. To highlight this aspect we performed another simulation in which we assumed the presence of only a $4^\textrm{th}$ order phase term. This situation could mimic the case in which the second order (GDD) term has been pre-compensated by a prism compressor. The results, for a $5.3\,\textrm{fs}$ laser pulse centered at $\omega_0 = 2.4 \, \textrm{rad/fs}$ are reported in Fig.\ref{fig:FourthOrder}. It can be seen that in this specific situation, not only Gated MIIPS better estimates the GDD but that standard MIIPS even lead to a reduction of the SHG intensity. Since pre-compensating the GDD before pulse-shaping is common practice, this result underlines the usefulness of Gated-MIIPS.

\FloatBarrier 

\section{Experimental Data}
\label{sec:ExpData}

\begin{figure}[htb] 
\centering\includegraphics[width=8.4cm]{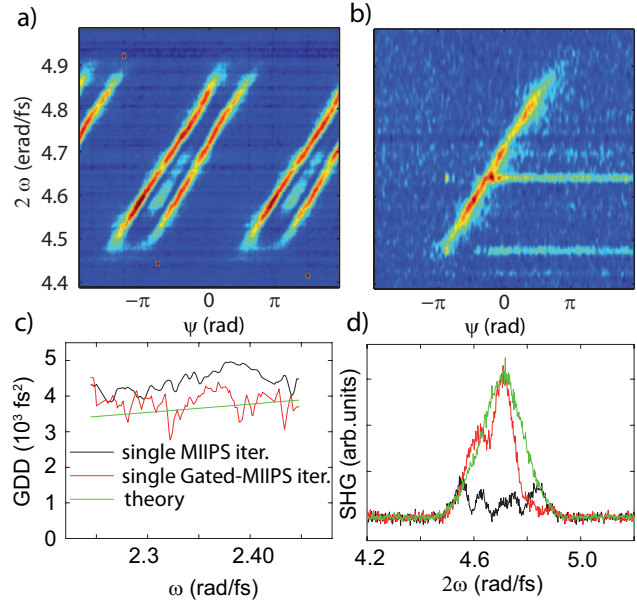}
\caption{Experimental comparison of MIIPS and Gated-MIIPS regarding the compensation of the  GDD introduced by $23\,\textrm{mm}$ of SF10 glass on a $15\,\textrm{fs}$ centered $800\,\textrm{nm}$. (a) standard MIIPS trace obtain with scanning parameter $\Phi_0 = 10\, {\rm rad}$ and $\tau = 25\, {\rm fs}$ (b) Gated-MIIPS iteration obtained with scanning parameter $\Phi_0 = 10\, {\rm rad}$, $\tau = 25\, {\rm fs}$ , $\sigma=0.18\,\textrm{rad}$. (c) GDD measured by a single MIIPS (black curve) and Gated-MIIPS (red curve) iteration, compared with the theoretical GDD calculated using the dispersion equation. d) SHG spectrum after a single iteration of either MIIPS (black curve) or Gated-MIIPS (red-curve), compared with the SHG obtained afer full pulse compression (green curve).}
\label{fig:GatedMiipsexperimen}
\end{figure}

In this section we experimentally compare the ability of MIIPS and Gated-MIIPS in estimating the chirp accumulated by a femtosecond laser pulse after traveling through glass. To perform the experiment, we started by compressing a $15fs$ laser pulse, centered at 800nm, and focused through a $1.3{\rm NA}$ microscope objective. Then we introduced additional glass on the optical path by a matched pair of SF10 prisms and performed a single iteration of either MIIPS or Gated-MIIPS. The two prims were in contact, with a thin layer of optical oil between them to minimize reflections at the interface.
 In both cases the scan parameters were: $\Phi_0 = 10 {\rm rad}$ and $\tau = 25 {\rm fs}$, with 401 phase steps and one second integration per step. The gate width was set to $0.18\,\textrm{rad}$. The measurements were performed using a 4f pulse shaping setup, based on a 128 pixel spatial light modulator. The SHG was obtained using BBO crystals, deposited on a microscope slide.

 The results, shown in Fig.\ref{fig:GatedMiipsexperimen}a, for $23{\rm mm}$ glass illustrate that the GDD introduced by the glass is better estimated by Gated-MIIPS. The contrast is particularly striking if one compare the resulting SHG signal after one iteration (\ref{fig:GatedMiipsexperimen}b): Gated-MIIPS in just one iteration recovers almost all the SHG spectrum of the compressed pulse. This result, together with the numerical simulations reported in the previous section, unambiguously show that Gated-MIIPS is a significant improvement towards a quick and reliable characterization of ultrashort laser pulses.

\section{Conclusions}
\label{sec:conclusions}

We introduced an improved scheme for ultrashort laser pulse compression and characterization,  which avoids some limitations of MIIPS for severely phase-distorted ultrashort pulses. We demonstrated its effectiveness in giving a more accurate correction to substantial phase distortions, as those caused by propagation in high-NA microscope objectives.



%

\end{document}